\documentclass[11pt,a4paper]{article}
\usepackage{jheppub}

\title{Holographic Entanglement Entropy and Confinement}

\author{Aitor Lewkowycz}

\affiliation{Perimeter Institute for Theoretical Physics,

Waterloo, Ontario N2L 2Y5, Canada}

\emailAdd{alewkowycz@pitp.ca}

\abstract{We study the phase transition in the holographic entanglement entropy for various confining models. This transition occurs for the entanglement entropy of a strip at a critical value of the strip width. Our main interest is to examine the critical width for models with several parameters.  For these models, the critical width, the glueball mass and the string tension all become functions of these two parameters. Comparing the behavior of the critical width in the entanglement entropy and these other scales, we find that 
$l_c$ seems to follow closely the deconfinement temperature and the glueball mass. The behavior of the string tension is similar to $l_c$, despite of being parametrically smaller than the other quantities. }

\begin{document}
\maketitle

\section{Introduction}
Entanglement entropy (EE) $S_A$ is a measure of entanglement between two subsystems $A,B$ of a quantum system (for a holographically centered review see \cite{Nishioka:2009un}). It is the von Neumann entropy of the reduced density matrix, obtained from the density matrix constructed by tracing out the degrees of freedom of our subsystem $B$  ($\rho_A=  \text{tr}_B  \rho \nonumber$):
\begin{align}
S_A=& -\text{tr}_A \left (\rho_A \log \rho_A \right ) \label{eq:entanglement}
\end{align}
 This quantity is $0$ if two subsystems are independent (ie. if the reduced density matrix still corresponds to a pure state). One can intuitively think of $S_A$ as the entropy due to lack of information of an observer in $A$ who can not access degrees of freedom in $B$. 

This definition of entanglement entropy is easily extended to quantum field theories as follows \cite{Calabrese:2004eu}: if the spacetime dimension of the theory is $d+1$ we first need to pick a $d$ dimensional time slice or Cauchy surface. Then we define a subsystem A to be a (spatial) subregion of this Cauchy surface and denote its complement $B$. The $d-1$ dimensional hypersurface which divides these two regions (the boundary of $A$, $ \partial A$)  is what we will call the entangling surface. With these considerations, we can define the entanglement entropy of the system as in (\ref{eq:entanglement}). Unfortunately, in a QFT, the result will always be divergent and we will have to introduce a cutoff $\xi$. The first divergent term is (for theories with $d>1$) proportional to the area ($\mathcal{A}$) of the boundary of $A$  :
\begin{align}
S_A \propto \dfrac{\mathcal{A}(\partial A)}{\xi^{d-1}}
\end{align} 
This relation is also called ``the area law''. However, the coefficient of this term is not universal, i.e. it depends on the way we choose the cutoff. After this term there can be sub-leading divergent terms (including a logarithmic term) and a finite constant term. The coefficient in front of the logarithmic term is universal and for CFT's with $d$ even, it is a combination of the central charges appearing in the trace anomaly. This relation between central charge and EE has been studied recently in \cite{Casini:2006es,Myers:2010xs,Myers:2010tj,Myers:2012ed}. 
The constant term has important uses in the condensed matter community \cite{Kitaev:2005dm,Levin:2006zz}. For topologically ordered systems, where there is no notion of order parameter, this quantity has proven to be the observable that characterizes the phase of the system.

The gauge/gravity duality allows us to perform calculations for strongly coupled $d+1$ dimensional field theories using holographic ($d+2$ dimensional) gravitational duals   \cite{Aharony:1999ti,Gubser:1998bc,Witten:1998qj}. Ryu and Takayanagi \cite{Ryu:2006bv,Ryu:2006ef} proposed a way to compute the entanglement entropy using holography.  The holographic entanglement entropy is obtained by considering the area of the minimal surface (extended in the extra direction) whose boundary ends in $\partial A$\footnote{Technically speaking, it is a saddle point, the area will decrease if we extend it in the time direction. However if we first Wick rotate to a Euclidean version of the theory, it is indeed minimal.}. If we denote this surface $\gamma$ (so that $\partial \gamma = \partial A$), the entanglement entropy will be given by \cite{Nishioka:2009un}  :
\begin{equation}
S_A=\dfrac{1}{4 G_N^{d+2}} \min_{\partial \gamma \sim \partial A }\mathcal{A}(\gamma) \label{eq:hee}
\end{equation}
This formula has not been proved in general but has been successfully verified with a wide range of consistency checks- see for example \cite{Headrick:2007km,Hung:2011xb}. Furthermore, the result was recently derived for the special case of a spherical entangling surface \cite{Casini:2011kv}. Although this definition of holographic entanglement entropy was first used for CFTs, we will use the same definition for studying nonconformal theories (introducing scales such as a confinement length or a flux).  It is worth 
commenting that there may be more than one extremal surface from which one must only consider the one with the minimal area. The absolute minima can be different if we change parameters of our geometry, so this defines naturally a phase transition in the EE, which occurs when two different surfaces have the same minimal area.

Ref \cite{Klebanov:2007ws} studied the behavior of the EE of a strip in holographic confining theories. On the gravity side these theories are realized by a compact cycle that shrinks to zero at some finite depth in the bulk. The holographic dictionary tells us that the bulk direction represents an energy scale and hence the minimal radii where the geometry ends introduces a mass gap in the dual theory \cite{Witten:1998zw}.  As we alluded to above,  (\ref{eq:hee}) can have
more than one competing extrema for confining theories. We will obtain a ``connected solution'' where the bulk surface runs between the two boundaries of the strip (and its area will depend on the width of the strip), but we will always also have a ``disconnected'' solution which has two independent pieces which go straight from the boundary to the cut-off radius (and whose area doesn't depend on the width of the strip). By comparing the areas of the connected and disconnected surfaces we will find that
they coincide for a particular width of the strip $l_c$. That means that if $l<l_c$ the EE will be obtained from the connected solution and for $l>l_c$ from the disconnected. In \cite{Klebanov:2007ws}, they then compute the value of $l_c$ for different systems and obtain that $l_c = O(1) \Lambda^{-1}_{IR}$. That is, the critical width seems to be closely related to the confinement scale. The phase transition in the EE for confining theories was also studied for different entangling geometries in \cite{Pakman:2008ui} with a similar result. This phase transition in the EE has been studied in different situations \cite{Faraggi:2007fu,Bah:2008cj,Fujita:2011fn,Cai:2012sk}. Setting holography asides,  this phase transition in the EE has also been found in numerical lattice simulations of $SU(N)$ gauge theories \cite{Buividovich:2008kq,Velytsky:2008rs}.

Motivated by \cite{Klebanov:2007ws}, in this note we would like to study what is the relation between the phase transition of the EE (connected-disconnected surfaces) and an underlying confinement- deconfinement phase transition with multiple parameters. 
In the holographic models studied in \cite{Klebanov:2007ws}, there is a single scale and so it is natural to expect that $l_c \sim 1/\Lambda_{IR}$ as well as $l_c \sim 1/T_c$ (where $T_c$ denotes the deconfining temperature). Our purpose is to test these relations in a more general context where the holographic models and hence the dual confining theory contains multiple parameters.
In this extension, we consider holographic models with one extra parameter which may help us understand better this transition. 
The EE phase transition will be determined by a critical width $l_c$, that is, the width of the strip in the boundary when the areas of the two contributions are the same. What we will do then is to study the behavior of this width and compare it with that of physical quantities that characterize our phase transition. 

The structure of the paper is as follows: In section 2, we set the preliminaries for the further calculations, that is, we review the computation of the EE for the strip and establish the procedure that we are going to follow in the next section. We also explain which 
quantities we are going to consider to compare with $l_c$. In section 3 we consider explicit examples with more than one scale where we study what is the dependence of $l_c$ on the extra scale, comparing its behavior with that of the other physical quantities presented in the previous section.
The models that we study can be constructed as the double Wick rotation of black hole solutions; the latter would be a charged $AdS_5$ BH, $D4$-$D0$ bound state and the backreaction of a relevant operator in the perturbative regime. We close the paper by discussing the results obtained.

\section{Preliminaries}

In this section we will review how to compute the minimal area that will give us the EE for our different holographic systems, for the entangling geometry of the strip.

The field theories we are going to study will be in a confined state, obtained from the finite temperature theories via a double Wick rotation, resulting in a solitonic solution. Our field theories will be $d+1$ dimensional with a compactified internal dimension whose proper length shrinks to zero in the bulk ($x_{d} \sim x_{d} + 2 \pi R $). A general metric with the desired characteristics will be (we are considering that the boundary is at $r=\infty$):
\begin{equation}
ds^2=-g_{tt}(-dt^2+d \vec{x}_{d-1}^2+f dx_{d}^2) + g_{rr} dr^2+ds'^2_{X} \label{eq:ansatz}
\end{equation}
We have included the possible presence of an internal manifold $X$, that usually appears in 10d supergravity solutions (but only appear as internal symmetries in the field theories). Where the compact field theoretical direction closes if we get deep in the bulk ($f(r_0)=0$). We have also included the possible presence of internal dimensions.

If we are dealing with a 10d supergravity solution with nonvanishing dilaton, the prescription for the EE is easier to understand with the Einstein frame, so we have to rescale our induced string frame metric such that:
\begin{equation}
 \sqrt{g^E_{ind}}=e^{-2 \phi} \sqrt {g_{ind}}
\end{equation}
Note that we will be integrating over 8 dimensions because if the gravitational theory is $d+2$ dimensional, our integral will be $d$ dimensional.

Given a metric with rotational symmetry in the spatial directions (as in our case for the $d-1$ spatial directions), the computation of the EE is straightforward for certain cases. In our calculations we will consider the region $A$ to be a strip and so the entangling surface are two flat planes separated by a distance $l$. Let's consider a strip of size $D$ in the $x_2,...,x_{d-1}$ directions and $l<<D$ in the $x_1$ direction, it is clear that (for the minimal surface embracing the strip in the boundary) only $x_1$ will depend on the $r$ coordinate. From these considerations it is easy to write the integral for the area:

\begin{align}
\mathcal{A}= & \int_X \int d^d x e^{-2 \phi} \sqrt{g_{ind}} =   D^{d-2}  2\pi R  \int_X \int_{-l/2}^{l/2} dx_1 e^{-2 \phi} \sqrt{g_{X}} f^{1/2} |g_{tt}|^{d/2} \sqrt{1+\dfrac{g_{rr}}{|g_{tt}|} \dot{r}^2} = 
\nonumber\\ = &D^{d-2}  2\pi R \text{Vol}(X) \int_{-l/2}^{l/2} dx_1 h_1 \sqrt{1+ h_2^2 \dot{r}^2} \label{eq:gind} 
\end{align}
In this equation $\dot{r} \equiv \frac{\partial r}{\partial x_1}$. In the last step, we factorized the volume of the internal manifold $X$ and just consider the integral of functions of $r$, remembering that if the dilaton
is nonzero we have to include it in $h_1$. Our task now is to find the minimal surface, so we will only have to minimize the integral. 

Because the integral doesn't depend on $x_1$ we can use ordinary functional methods to argue that the ``energy'' is conserved  ($E=\dfrac{\partial \mathcal{L}}{\partial \dot{r}} \dot{r}- \mathcal{L}$) and then obtain the maximum depth in the $r$ direction of our surface, if we denote this value of the coordinate $r^{\star}$ ($h^{\star} \equiv h(r^{\star}) $), we can rearrange our integral in one that we know how to compute given the maximum depth:

\begin{align}
 & \dot{r}=h_2^{-1}  \sqrt{\left ( \dfrac{h_1}{h_1^{\star}} \right ) ^2  -1} \nonumber \\
 & \dfrac{l}{2}=\int_{r^{\star}}^{\infty} \dfrac{d r} {\dot{r}} \nonumber  \\
& \mathcal{A}^c = D^{d-2}  2\pi R \text{Vol}(X) \int_{r^{\star}}^{\infty} dr \frac{h_1 h_2}{\sqrt{1 -\left ( \dfrac{h_1^{\star}}{h_1} \right ) ^2  }}=\int_{r^{\star}}^{\infty} dr \mathcal{F}^c \label{eq:eestrip}
\end{align}

This was the computation for the connected surface, there is also a disconnected surface (which does not depend on $l$), that can be easily obtained from (\ref{eq:gind}) by demanding $\dot{r}^{-1}=0$:
\begin{equation}
 \mathcal{A}^d=   D^{d-2}  2\pi R \text{Vol}(X) \int_{r_0}^{\infty} d r h_1  h_2 =\int_{r_{0}}^{\infty} dr \mathcal{F}^d
\end{equation}

Now we are interested in compute the difference in EE between the connected and disconnected pieces. 
Because the two areas diverge in the same way (as the integral is extended to the boundary), we can get the difference in the finite piece just by subtracting inside the integral:

\begin{equation}
\Delta \mathcal{A}(r^{\star})=\int_{r^{\star}}^{\infty} (\mathcal{F}^c-\mathcal{F}^d)-\int_{r_0}^{r^{\star}} \mathcal{F}^d \label{eq:eeint}
\end{equation}

In this way, we will define $l_c$ as the width where the disconnected and connected piece have the same areas:

\begin{equation}
\Delta \mathcal{A}(r_c)=0 \rightarrow l_c=l(r_c) \label{eq:lc}
\end{equation}

\subsection{Physical scales in confining theories}

The physical scales that we will consider for the confining models will be the deconfinement temperature ($T_c$), the glueball mass ($M_{gb}$) and the tension of the flux string ($T_{str}$). We are interested in comparing these energy scales to $l^{-1}_c$.

\begin{itemize}
 \item Deconfinement temperature:

We will have two gravitational backgrounds, corresponding to the confining and deconfining phases - both of them are periodic in (at least) one spatial direction and the imaginary time direction. Their Euclidean action will depend on the radius of confinement ($R$), the temperature ($T$) and the external parameter ($Y$). 
Following the standard approach of Euclidean quantum gravity, the phase with the largest (negative) action dominates. Hence equating the gravitational actions at the same $R,T$ will give us the deconfinement temperature as a function of $R$ and $Y$:
\begin{equation}
 \Delta I (R,T_c,Y)=0 \rightarrow T_c(Y,R) 
\end{equation}

\item Glueball mass:

We can compute the glueball mass of our theory by considering a massless scalar field propagating in a background, see for example \cite{Csaki:1998qr,Minahan:1998tm,Constable:1999gb}. Using the ansatz: $\phi=\psi(r) \beta(r) e^{-i M t}$, the scalar dynamics are determined by the wave equation:
\begin{align}
\partial_{\mu} \left ( \sqrt{-g} g^{\mu \nu}\partial_{\nu} \phi \right )= & 0 
\end{align}
%
This wave equation can be easily put into a Schrodinger form, but if we do the crudest WKB approximation to get 
a feeling of how the mass depends on the parameters of our theory, we don't really need to do the algebra. In this way, if we consider that the glueball is very massive so we neglect other terms proportional to $\psi(r)$, for large $n$, WKB yields \cite{Minahan:1998tm}:

\begin{equation}
 M_{gb}^{-1} \equiv M^{-1} (n-\frac{1}{2})  = \frac{1}{\pi} \int_{r} \sqrt{\dfrac{-g^{tt}}{g^{rr}}} dr \label{eq:glueballmass} 
\end{equation}

The integral is definite and we have to integrate $r$ from the boundary to the end of the geometry. Although $M$ is not the first excitation, it gives us a scale for the glueball mass spectrum and because we only worry about what is the scale that the glueball is defining, we will reabsorb the $n$ factors in this energy scale ($M_{gb}$). Abusing of language, we will call $M_{gb}$ the glueball mass, even though 
it is just the scale of the large excitations of the glueball spectra (for large $n$, that we reabsorbed in its definition). 

\item Tension of the string:

The tension of the string is defined for the confining solution as the effective fundamental string tension redshifted at $r_0$ \cite{Witten:1998zw}:

\begin{equation}
 T_{str} = \frac{1}{2 \pi \alpha'} \sqrt{-g_{xx} g_{tt}} \big |_{r=r_0} \label{eq:tension}
\end{equation}

 Note however that this quantity will be parametrically different than the previous ones because it usually comes multiplied by the 't Hooft coupling $\lambda$, and the calculation we will make will be in the strong coupling regime.

\end{itemize}

\section{Different models}

\subsection{5D flux soliton}

The model is a magnetically charged extension of the first case considered in \cite{Klebanov:2007ws}. We may begin with the charged black hole in $AdS_5$, which is a solution to the Einstein-Maxwell equations of motion and can be uplifted to type IIB SUGRA (see  \cite{Chamblin:1999tk}). Wick rotating this solution we get a ``magnetically charged'' soliton. This model has two scales, the potential $\Phi$ at infinity and the compactification radius of the spatial direction $x_3$.
In order to study a confinement/deconfinement phase transition, we will construct two solutions from this black hole: a soliton with magnetic flux in a compactified coordinate and a black hole with constant magnetic flux in the compact direction.
We expect that if there is a phase transition the confinement phase will be described by the thermal soliton and the black hole will represent the deconfined phase. 

As noted, given the solution of the charged black hole, we will do a double Wick rotation to get the soliton (in order to keep the potential real, we will reabsorb a imaginary unit in the potential) and we will assume periodic imaginary time. The other solution will simply be the black hole with constant potential.
 In this way, for the two cases, the coordinates $t, x_3$ should be identified as $t \rightarrow t+ i T^{-1}, x_3 \rightarrow x_3 + 2 \pi R$. The solutions we will be considering are then  :

\begin{align}
 \text{Soliton: } & ds^2_c=  \dfrac{r^2}{L^2} \left[-dt^2+dx_1^2+dx_2^2+f_c dx_3^2 \right  ]+\dfrac{L^2 d r^2} {r^2 f_c}  \nonumber \\
& A_c= \sqrt{3} \Phi (1-\frac{r_0^2}{r^2}) dx_3 \hspace*{6mm} f_c=\left  (1-\frac{r_0^2}{r^2} \right) \left ( 1+\frac{r_0^2}{r^2}+\frac{\Phi^2 r_0^2 L^2}{r^4} \right ) \nonumber \\
 \text{Black Hole: } & ds^2_d= \dfrac{r^2}{L^2} \left[-f_d dt^2+dx_1^2+dx_2^2+ dx_3^2 \right  ]+\dfrac{L^2 d r^2} {r^2 f_d}  \nonumber \\
& A_d=  \sqrt{3} \Phi dx_3  \hspace*{19mm}  f_d= \left  (1-\frac{r_d^4}{r^4} \right) \label{eq:31}
\end{align}

The subscripts $c,d$ which denote confined and deconfined phase respectively and correspond to the soliton and black hole. Note that, because the $x_3$ direction is compact, the gauge potential in the deconfined phase is not pure gauge, we have a nontrivial magnetic holonomy.
From here it is easy to compute the radius of the compact direction and the temperature of the black hole in terms of the parameters of the theory (they have to be fixed in order to avoid conical singularities).

The radius of the compact direction $x_3$ is:
\begin{align}
 \frac{1}{R}=& \dfrac{r_0}{L^2} \left ( 2+\dfrac{\Phi L}{r_0} \right )  \nonumber \\
 r^{\pm}_0=& \frac{L^2}{4 R}(1 \pm \sqrt{1-8 x^2})  \hspace*{0.4cm} \text{with }   x=\dfrac{\Phi R}{L}
\end{align}

Like we said before, there are only two independent parameters (we also introduce a temperature implicitly but $r_0$ is independent of $T$). The ones that are physically interesting are $R$ and $\Phi$, and we have seen that $r_0(R,\Phi)$ has two solutions for every pair of parameters. However, because we know that at $x=0$, $r_0=\frac{L^2}{2 R^2}$,
the only physical solution is $r^{+}_0$, so from now on we will omit the $\pm$ label. Also, because we want $r_0$ to be real, the variable $x$ will have to be between $0$ and $\frac{1}{\sqrt{8}}$. 
%

The temperature of our system will be determined by identifying periodically imaginary time for the black hole:
\begin{equation}
 T^{-1}=\frac{\pi L^2}{r_d}
\end{equation}

Although we have two solutions at finite temperature for the same system, at every point $T,R,\Phi$ the state will be described by one of the solutions. The physical solution is the one which has the least (i.e. more negative) free energy. The difference between the free energy of the two solutions goes like:
\begin{equation}
 \Delta I_{c-d} \propto R T^{-1} \left [ r_d^4-r_0^4-(\Phi L r_0)^2\right ]
\end{equation}

If we now plug the expressions for $T$ and $R$ in terms of the parameters it is straightforward to find the deconfinement transition (if $T<T_c$ we are in the confined phase):

\begin{eqnarray}
T_c^4=\dfrac{1}{(2 \pi R)^4} h(x) \nonumber \\ 
h(x)=\frac{1}{2} (1-4 x^2 - 8 x^2 + \sqrt{1-8 x^2})
\end{eqnarray}

\begin{figure}
      \centering
    \includegraphics[width=0.5\textwidth]{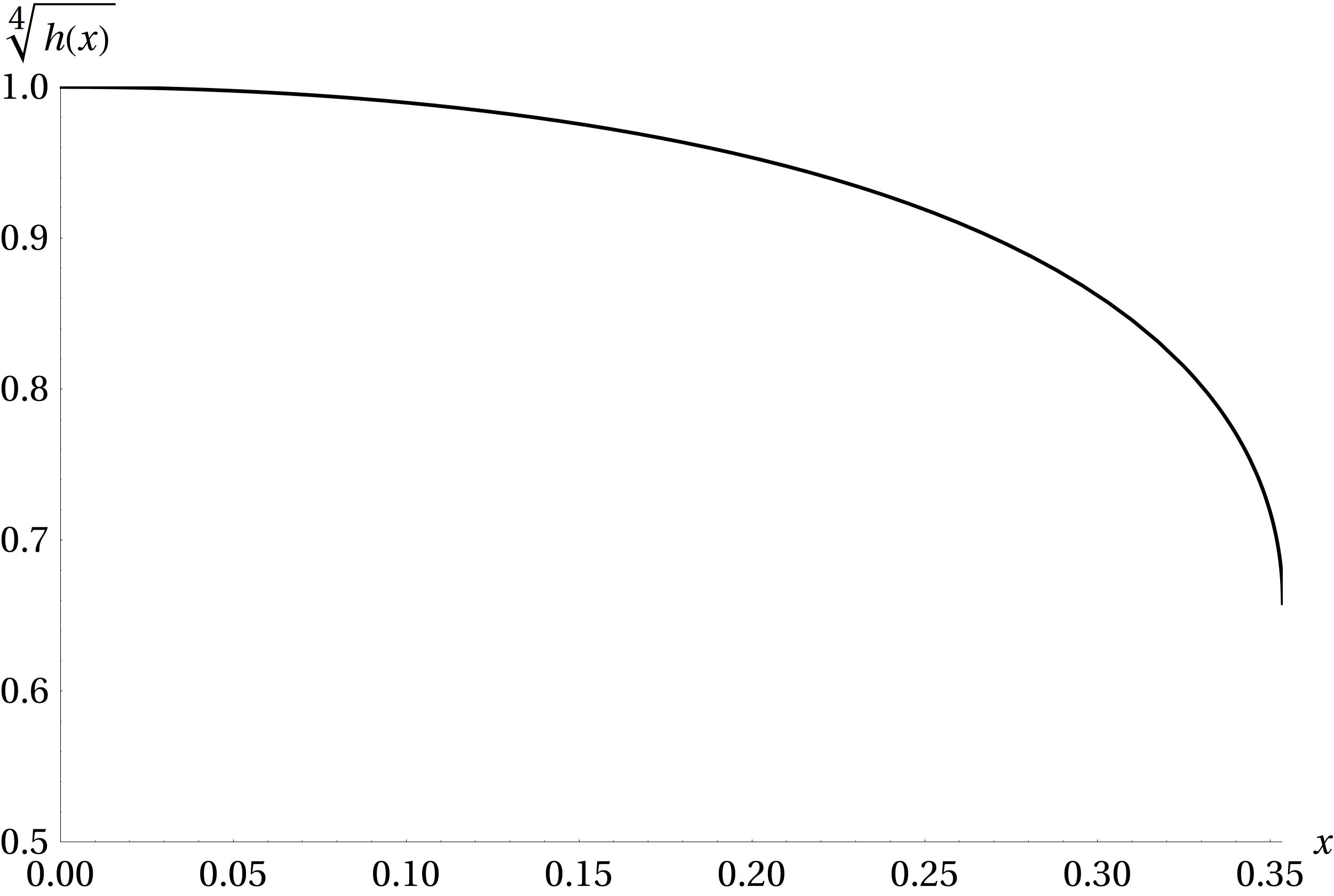}
  \caption{$h(x)^{1/4}=\dfrac{T_c}{2 \pi R} $ \label{fig:hx}} 
\end{figure}

We can see the function $h(x)^{1/4}$ plotted in figure \ref{fig:hx}, because $x\in [0,\frac{1}{\sqrt{8}}] \rightarrow h(x) \in [1,\frac{3}{16}]$.

The glueball mass \ref{eq:glueballmass}  and tension \ref{eq:tension} will be:

\begin{eqnarray}
 \pi M^{-1}_{gb}   \frac{1+\sqrt{1-8 x}}{4 R} =a_{gb}(x)= \int_{1}^{\infty} \dfrac{1}{\sqrt{(1-y^{-2})(1+y^{-2}+x^2 y^{-4})}} dy \\
T_{str} = \frac{1}{2 \pi \alpha'}  \left ( \frac{r_0}{L} \right )^2=  \frac{ \lambda}{16 R^2} (1+\sqrt{1-8 x})^2
\end{eqnarray}
where $\lambda$, the 't Hooft coupling is defined as usual $\lambda=\frac{L^2}{2 \pi \alpha'} $

\subsubsection{Stripe EE}

With all the previous things in consideration, the computation of the entanglement entropy is straightforward: everything will reduce to a relatively complicated integral of a function of $r$ between $r^{\star}$ and $\infty$ (\ref{eq:eestrip}) , where $r^{\star}$ is as usual how deep our surface gets in the $r$ direction, and after integrating another function we will obtain the physical width $l(r^{\star})$. In order to simplify the final expressions we did some change of variables. The first change of 
variables is to go to a coordinate where the boundary is at $0$ and the upper limit of the integral is $1$. This coordinate is: $y=\dfrac{r^{\star}}{r}$. We have a family of EE parametrized by $\Phi$ and $r^{\star}$ (which is an implicit function of $l$), to make the computations easier let's
assume  $r^{\star}=\frac{r_0}{\mu}$, the final redefinition that we will make is $ q \equiv \dfrac{\Phi L}{r_0} = \dfrac{4 x}{1+\sqrt{1-8 x^2}}$. In this way, we will be able to factorize the dependence in $r_0$ to end up having just a family of integrals parametrized by $\mu$ and $q$, whose values must be between $[0,1]$ and $[0,\sqrt{2}]$, respectively.

With all these simplifications, using the usual formula for the EE, the integral that we will have to compute to get the entanglement entropy is the following (we set $L=1$):

\begin{align}
 S (l,\Phi)  & =  \dfrac{ \pi R  D}{ G^{5}_N}  \mathcal{{A}}[l,\Phi]    \nonumber \\
\mathcal{A}[l,\Phi]& =r_0^{2}  \mathcal{\bar{A}}[l,q(\Phi)]=r_0^{2} \int_0^1 dy \hspace*{0.1cm}  \mathcal{F} [y,\mu(l),q] \nonumber \\
l(\mu,\Phi) & = \frac{\bar{l}(\mu,q(x))}{r_0(x)} =\frac{2}{r_0} \int_0^1 dy \frac{dl}{dy} \label{eq:EEcharged}
\end{align}

Note that we explicitly wrote the dependence of the expressions on $r_0$, this factorization will make the numerics much easier and comes precisely from the choice of coordinates we did. From now on we will drop the dependence on $x$ of $q$ and $r_0$ (we know their explicit dependence on $x,R$).
With this, we can integrate the functions $\mathcal{F}$ (note that the second piece of $\mathcal{F}^d$ in (\ref{eq:eeint}) has to be integrated between $1$ and $\mu^{-1}$)  and $\frac{dl}{dy}$ between $0,1$ for a set of values of $\mu$ and obtain the functions $\Delta \mathcal{\bar{A}}[\bar{l},q]$, the ones that are of interest. The functions to integrate are:

\begin{align} 
\mathcal{F}^c[y,\mu,q]  =& \sqrt{\frac{1-\left(1-q^2\right) y^4 \mu ^4-q^2 y^6 \mu ^6}{y^6 \mu ^4 \left(1-y^6+\left(1-q^2\right) y^4 \left(-1+y^2\right) \mu ^4\right)}} \label{eq:toint2}  \\
\mathcal{F}^d[y,\mu,q]  =&\frac{1}{y^3 \mu ^2}  \nonumber \\
\dfrac{dl}{dy}=&\sqrt{\frac{y^6 \mu ^2 \left(-1+\mu ^2\right) \left(-1-\mu ^2-q^2 \mu ^4\right)}{\left(-1+y^2\right) \left(-1+y^2 \mu ^2\right) \left(-1-y^2 \mu ^2-q^2 y^4 \mu ^4\right) \left(-1-y^2+y^4 \left(-1+\left(1-q^2\right) \mu ^4\right)\right)}}
 \nonumber
\end{align}

\begin{figure}
  
    \centering
    \includegraphics[width=0.5\textwidth]{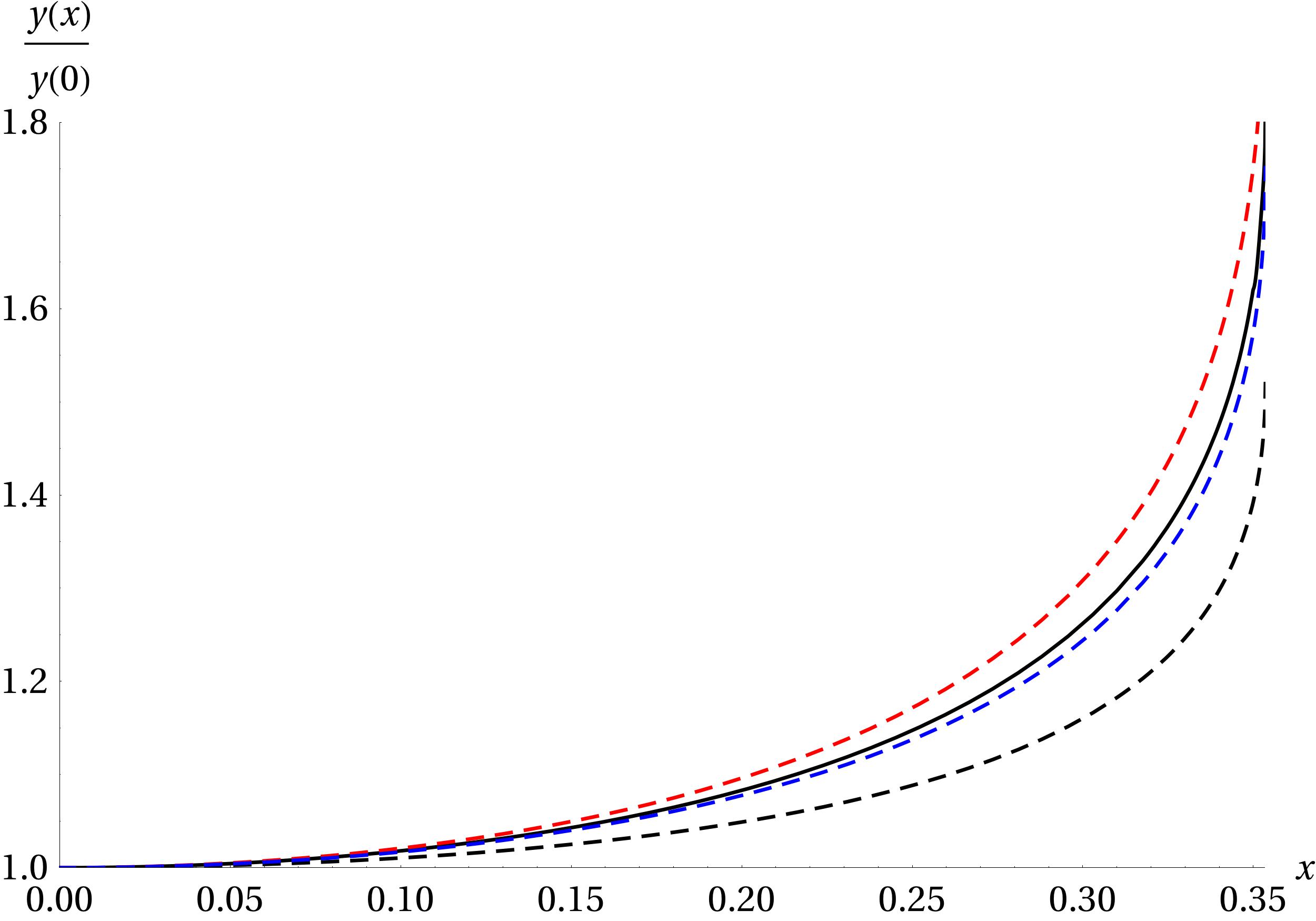}
  \caption{(Colour online) Dependence on $x$ for $l_c$ (black) and the scales of our theory 
(dashed) $1/T_c$ (black), $T_{str}^{-1/2} $ (red) and $M^{-1}_{gb} \propto a(x)/r_0$ (blue), 
the quantities are normalized with their value at $x = 
 0 $, so they all begin at 1. \label{fig:lccharged}}

\end{figure}

Integrating these functions we can obtain $l_c(q)$ by equating $\Delta \mathcal{\bar{A}}[l_c,q]=0$. With everything we have, now it is relatively easy to get this values. For every $x$ we will have to obtain the functions $\mu_c(x)$ by solving (\ref{eq:lc}) numerically. After that, we plug $\mu_c$ and $x$ in the integral for the physical width, so we
end up having the function $l_c(x)$. We can see these quantities plotted in figure \ref{fig:lccharged}. Note that although the functions may seem to be diverging, they are finite at $x=\frac{1}{\sqrt{8}}$.

 We can see how $l_c$ tracks well the other physical three quantities, in particular it seems to follow most closely the glueball mass. These results are not conclusive about the preferences of $l_c$, because it seems all of the physical quantities of interest to track to order one. Of course all of the relevant quantities track each other at the qualitative level.

\subsection{Magnetically charged D4 soliton}

In order to obtain this model we will begin with the soliton M5 brane supergravity solution (for example \cite{Ortin:2004ms}):
\begin{align}
 ds_{11}^2=H_3^{-1/3} (d \vec{x}^2+f dx_4^2+dx_{11}^2)+ H_3^{2/3} (\frac{dr^2}{f}+r^2 d \Omega_4 ^2 ) \nonumber \\
C_{t x_1 x_2 x_3 x_4 x_{11}}=\sqrt{1+\left (\frac{r_0}{L}\right)^3}(H_3^{-1}-1) \approx H_3^{-1} 
\end{align}
where:
\begin{eqnarray}
 H_3=1+\left (\frac{L}{r}\right)^3 \approx \left (\frac{L}{r}\right)^3  &  &  f=1-\left (\frac{r_0}{r} \right)^3
\end{eqnarray}
Here the parameter $L$ is specified in terms of the string theory parameters and the near horizon limit ($L>>r$) that we used ends up being the usual field theory limit (large $N_c$ but finite ``t'Hooft coupling''). If we rotate in the $4$ and $11$ direction with an angle $\theta$ we obtain the rotated solution:
\begin{align}
 \nonumber \\
ds_{11}^2= & H_3^{-1/3} (d \vec{x}^2+f H_0^{-1} dx_4^2)+ H_3^{2/3} (\frac{dr^2}{f}+r^2 d \Omega_4 ^2 )+H_3^{-1/3} H_0 \left [ dx_{11} + \text{cotan} \theta \left ( H_0^{-1}-1 \right ) dx_4 \right ]^2
\nonumber \\ H_0= & 1-\frac{q^3}{r^3}  \hspace{3mm} \text{with } q^3=r_0^3 \sin ^2  \theta 
\end{align}
In order to KK compactify the 11th dimension we will decompose it as usual:

\begin{equation}
 ds_{11}^2=e^{-2\phi/3} g_{\mu \nu} dx^{\mu} dx^{\nu}+e^{4 \phi/3} \left [ dx^{11}+ A_{\mu} dx^{\mu} \right ]^2
\end{equation}
So we end up with the result we wanted (we add a constant to the gauge potential to ensure $A(r_0)=0$):

\begin{equation}
 ds_{c}^2 =  H_3^{-1/2} \left [-f H_0^{-1/2} dx_4^2+ H_0^{1/2}(-dt^2+d{x^2_1}+d{x^2_2}+d{x^2_3}) \right] + H_0^{1/2} H_3^{1/2}  \left (\dfrac{dr^2}{f}+r^2 d \Omega_4 ^2 \right )  
\end{equation}
where the dilaton, RR 3-form and RR 1-form are:
\begin{align}
 e^{-2 \phi}=g_s^{-2} H_0^{-3/2} H_3^{1/2} \hspace{8mm} F_4= g_s^{-1} 3 L^3 d \Omega_4 \hspace{8mm} A=\dfrac{1}{\sqrt{\frac{r_0^3}{q^3}-1}} \dfrac{f}{H_0} dx_4 =\Phi \dfrac{f}{H_0} dx_4 
\end{align}

This solution has a magnetically charged RR 1-form, which is sourced by D6 branes, so it is natural to think of this solution as a D4-D6 brane bound state. Physically, the parameters we are interested in are the confining radius and the potential at infinity $\Phi$ (in terms of the angle of the rotation $\Phi=\tan \theta$).  We can easily compute the confining radius of this soliton:
\begin{equation}
 R=\frac{2}{g_{x_4 x_4}'(r_0)}=\frac{2}{3} r_0 \sqrt{H_0 H_3} = \frac{2 L^{3/2}}{3 r_0^{1/2}} (1+\Phi^2)^{-1/2}
\end{equation}

Because the physical quantities are $\Phi$ and $R$, we can express $r_0$ and $q$ in terms of these variables :

\begin{align}
r_0 & =\frac{4 L^3}{9 R^2} \frac{1}{1+\Phi^2} \nonumber \\
q &=\frac{4 L^3}{9 R^2}  \frac{1}{(1+\Phi^2)(1+\Phi^{-2})^{1/3}}=\frac{4 L^3}{9 R^2}  \frac{\Phi^{2/3}}{(1+\Phi^2)^{4/3}}
\end{align}

If we now consider the deconfined solution, it'll just be the black hole solution of the $D4$ background, with a constant gauge potential and the $x_4$ direction compactified in a circle.
It'll simply be then:
\begin{equation}
ds_{d}^2=H_3^{-1/2} \left [-f(r_d)  dt^2+ d{x^2_1}+d{x^2_2}+d{x^2_3}+d{x^2_4} \right] +  H_3^{1/2}  \left (\frac{dr^2}{f(r_d)}+r^2 d \Omega_4 ^2 \right )    
\end{equation}
Where the dilaton, RR 3-form and RR 1-form are:
\begin{align}
 e^{-2 \phi}= g_s^{-2}  H_3^{1/2} \hspace{18mm} F_4= g_s^{-1} 3 L^3 d \Omega_4  \hspace{18mm} A= \Phi dx_4
\end{align}
The temperature of this black hole is:
\begin{equation}
 T=\frac{3 r_d^{1/2}}{4 \pi L^{3/2}}
\end{equation}

In that way, if we are to compare the action of the two solutions, they have to have the same compact circles (ie. time periodic in imaginary time with period $T^{-1}$ and $x_4$ with period $2 \pi R$). 
{ If we compare these two solutions, from M theory the two solutions only differ by a rotation, and because nothing else than the metric depends on the rotation (which only acts through the determinant and the norm of $F_3$), the two actions are the same (with different parameters $r_0$, $r_d$). From this, the deconfinement temperature will be determined by $r_0=r_d$:
\begin{equation}
  T_c^{-1}=\frac{4 \pi L^{3/2}}{3 r_0^{1/2}}=2 \pi R \sqrt{1+\Phi^2}
\end{equation}

 }

The glueball mass and the tension of the confined string are:

\begin{eqnarray}
 M^{-1}_{gb} =\frac{1}{\pi} \int_{r_0}^{\infty} dr \sqrt{\dfrac{L^3}{r^3(1-(r_0/r)^3)}}= \frac{1}{\pi} \sqrt{\dfrac{L^3}{r_0}} \int_{1}^{\infty} dy \dfrac{1}{\sqrt{y^3(1-y^{-3})}} \nonumber \\
M_{gb} \simeq  5.41843  T_c \\
\nonumber \\ 
 T_{str} = \frac{1}{2 \pi \alpha'}  H_0^{1/2} H_3^{-1/2} \big |_{r=r_0} =\frac{1}{2 \pi \alpha'} \frac{3 R }{2 L^3} r_0^2 \sim  \lambda R T_c^4 
\end{eqnarray}

Where in the last line we got rid of all the numerical factors, and we subsituted the t' Hooft coupling in 5d $\lambda \sim \frac{L^3}{2 \pi \alpha'}$ (it has length dimensions). Note that in this case, for $\Phi=0$, $T_{str}^{-1/2} \sim \sqrt{R^3 \lambda^{-1}}$, so although the power of $R$ is not the same that the one for $l_c$, we are still comparing $l_c$ with $T_{str}^{-1/2}$ because it is a fundamental quantity in the confining theory with length units and its behavior with $q$ could be similar.
\subsubsection{Stripe EE}

Now consider the EE for a strip, if we change to coordinates $z=1/r$ and we consider that our strip is a 8 dimensional surface that is completely characterized by the profile $z(x)$, the induced metric is:

\begin{equation}
 \sqrt{g^E_{ind}}=\frac{f^{1/2} H_3^{1/2}}{z^4} \sqrt{1+\frac{H_3 z'^2}{f z^4}}
\end{equation}

Because there is no explicit dependence on $H_0$,  the problem is the same one that without the $D0$ branes but with different identifications ($r_0$ also depends on $\Phi$).

If we are to compare the two competing extremal surfaces, we will find that the critical width is:

\begin{equation}
 l_c^{-1} \simeq 1.1647  \sqrt{\frac{r_0}{L^3}} =  4.8786 T_c
\end{equation}

 In this system both $l_c$ and $M^{-1}_{gb}$ track  $T_c$ (note also that $l_c M_{gb} \sim 1$), while $T_{str}^{1/2}$ goes like $T_c^{2}$. An interesting observation about this case is that the flux can be as large as we want, so we can completely separate $1/T_c$ from $R$. In this way what the EE phase transition
seems to tell us is that $l_c$ goes like $M^{-1}_{gb}$ or $T_c^{-1}$.

\subsection{Relevant perturbation}

 In this section we will consider our thermal theory to be deformed by a relevant perturbation. For simplicity we will limit ourselves to first order in the perturbation (i.e. we work with large ``temperature''), which will allow us to have analytic functions until relatively far. We are going to consider as usual a confinement-deconfinement transition.
We will follow the notation of \cite{Buchel:2003ah,Buchel:2004hw,Buchel:2007vy} and most of the results here included (before the EE) can be found there.

The Einstein-Hilbert action that we are considering is (we set the radius of the asymptotic $AdS_5$ to be $L=2$):
\begin{eqnarray}
 S=\frac{1}{16 \pi G_5} \int_{\mathcal{M}_5} d \xi ^5 \sqrt{-g} \left ( R - 4 (\partial \phi)^2 -V(\phi) \right ) \nonumber \\
V(\phi)= -3 + 4 m^2 \phi^2
\end{eqnarray}

Note that the normalization of the scalar is not the canonical one (which will be just $4 \phi^2 \rightarrow \frac{1}{2} \phi^2 $).  $m$ denotes mass of the scalar and is related to the dimension of the operator inserted in the boundary in the usual manner \cite{Aharony:1999ti,Witten:1998qj}: $m L=\sqrt{\Delta (\Delta-4)}$. The physical dimensions we consider for the operator are $\Delta \in (2,4)$ (note that then $m^2 \le 0 , |m^2| <1 $).

In the regime of small $\phi$  (its smallness is controlled by $\phi_0$), we can solve the equations of motion analytically.  The metrics\footnote{We again follow  the conventions of \cite{Buchel:2007vy}, comparing with the previous ones (\ref{eq:ansatz}), $(1-x)^2=-f g_{tt}$, $c_2=\sqrt{-g_{tt}}$, $G_{xx}=g_{rr} \left ( \frac{dr}{dx} \right ) ^2$} (which are basically \ref{eq:31} with $\Phi=0$ and the backreaction of a scalar and, again, we periodically identify $t,x_3$:$t \rightarrow t+i T^{-1}, x_3 \rightarrow x_3 + 2 \pi R$ ) and scalar profile are:

\begin{eqnarray}
\phi(x)= \phi_0 (2x-x^2)^{\Delta/4} {}_1F_2 \left (\dfrac{\Delta}{4},\dfrac{\Delta}{4};1;(1-x)^2 \right ) \nonumber  \\ 
ds_c^2=  c_2^2 (-dt^2+dx_1^2+dx_2^2+(1-x)^2 dx_3^2)+G_{xx} d x^2 \nonumber \\
ds_d^2=c_2^2 (-(1-x)^2 dt^2+dx_1^2+dx_2^2+ dx_3^2)+G_{xx} d x^2 \nonumber \\
c_2= \dfrac{\delta_{c/d}}{(2x-x^2)^{1/4}} e^{A(x)} 
\end{eqnarray}

Here $A(x)$ and  $G_{xx}$ (see \cite{Buchel:2007vy})  can be easily obtained from the equations of motion ($G_{xx}$ is expanded to leading order in $\phi_0^2$):

\begin{eqnarray*}
 A(x)=\frac{4}{3} \int_x^1 \frac{(z-1)dz}{(2z-z^2)^2}(B(0)-B(z)) \\
 B(z)=\int_z^1 dy \phi'(y)^2 \frac{(2y-y^2)^2}{(y-1)} \\
 G_{xx}=\frac{1}{(2x-x^2)^2}-\frac{2A'}{2x-x^2} \left ( (1-x)+(1-x)^{-1} \right )+\frac{4 m^2 \phi ^2}{3 (2x-x^2)^2}-\frac{4}{3} \phi'^2 
\end{eqnarray*}

For $A(x)$ we have the freedom to add a constant to $B(z)$ and we choose it so the integral is finite at $y=0$.

Like in the previous cases, $t$ and $x_3$ are identified as $t \rightarrow t+i T^{-1}, x_3 \rightarrow x_3 + 2 \pi R$. In order to compare the two theories we will suppose they have the 
same $\phi_0,T,R$. In this way, from the usual definition for temperature of a gravitational system ($2 \pi T= c_2 G_{xx}^{-1/2} \big |_{x=1}$),  $T$ and $R$ will be \cite{Buchel:2007vy} :
\begin{equation}
 \dfrac{1}{R \delta_c}=\dfrac{2 \pi T}{ \delta_d}= 1 + \left ( \frac{\Delta (4-\Delta)}{6}-A''(x) \big |_{x=1} \right ) \phi_0^2= 1+c_R \phi_0^2
\end{equation}

Note that in general $\delta_c \not = \delta_{d}$. Because the deformation is the same in both backgrounds, the transition will occur at $T^{-1}= 2 \pi R$ (or $\delta_c=\delta_d$). 

The tension of the string and the glueball mass will be:

\begin{align}
  T_{str} = \frac{1}{2 \pi \alpha'} \sqrt{-g_{x_1 x_1} g_{tt}} \big |_{x=1}=\frac{1}{2 \pi \alpha'} c_2^2|_{x=1}=\frac{\lambda}{4} \frac{1}{(1+c_R \Phi_0^2)^2 R^2} \nonumber \\
 M_{gb}^{-1} = \frac{1}{\pi} \int_0^1 dx \sqrt{\dfrac{G_{xx}}{c_2^2}}= \frac{1}{\pi} \delta_c^{-1} a_{gb}(\phi_0)
\end{align}

Because we only calculate up to first order in $\phi_0^2$, for every physical variable,the quantities that we will be interested in will be:

\begin{equation}
 \frac{X(\phi_0)}{X(0)}=1+c_X \phi_0^2
\end{equation}

For the two quantities before then:

\begin{eqnarray}
 c_{\sqrt{T_{str}}}=&-c_R \nonumber \\ 
 c_{M^{-1}_{gb}}=&c_R+\dfrac{a_{\Delta}}{a_0} &; a_{gb}=a_0+a_{\Delta} \phi_0^2 
\end{eqnarray}
where in the second case, we expanded $a_{gb}$ in the perturbation. In this way, in order to get $c_{M^{-1}_{gb}}$, we will only have to do the integral $a_{\Delta}$ for every $\Delta$.

\subsubsection{Stripe EE}

What we want to do now is to compute $l_c$, the width at which the disconnected piece of the entanglement entropy has the same area
than the connected one. First we construct the area functionals for the two cases:

\begin{eqnarray}
 \mathcal{F}^c=\dfrac{ (1-x) c_2^2 G_{xx}^{1/2}}{\sqrt{1-\left (\dfrac{(1-x^{\star})c^{\star 3}_2 }{ (1-x)c_2^3} \right ) ^2}} \nonumber\\
 \frac{dl}{d x}=\frac{G_{xx}^{1/2}}{c_2 {\sqrt{\left (\dfrac{(1-x)c_2^3 }{(1-x^{\star}) c^{\star 3}_2 } \right ) ^2-1}}}  \nonumber \\ 
 \mathcal{F}^d=(1-x) c_2^2  G_{xx}^{1/2}
\end{eqnarray}

In the usual way, the $\star$ denotes the tip of the strip, which will be the upper limit of our integration and we can obtain the physical width by 
integrating  $ \dot{l}$. Because we are working in the small $\phi_0$ limit, we can only trust the results up the first quadratic term in $\phi_0$, which will make everything simpler.
We will also have to expand then in $x^{\star}=x_0+\phi_0^2 \delta x$:
\begin{eqnarray}
 \mathcal{F}=\delta_c (\mathcal{F}_0+\phi_0^2 \mathcal{F}_{\Delta}+\phi_0^2 \mathcal{F}_{\delta x} \delta x ) \nonumber \\
 \frac{dl}{d x}=\delta_c^{-1} (\dot{l}_0+\phi_0^2 \dot{l}_{\Delta}+\phi_0^2 \dot{l}_{\delta x} \delta x )
\end{eqnarray}

Here we explicitly extracted the dependence in $\delta_R$ so all the functions $\mathcal{F}_i, \dot{l}_i$ only depend on $x,x_0$.The shift of $x^{\star}$ due to the perturbation is linear and the only term that depends explicitly on the dimension of the operator is the one with subscript $\Delta$ (the last term comes only from expanding in $x^{\star}$).  We are interested in the difference of the constant piece of the 
EE for the two solutions and, because the divergent piece will be the same, the integral to compute will be the following (note that for the term linear in $\delta x$ there is no disconnected piece and it is convergent):
\begin{equation}
 \mathcal{A}_i(x^{\star})=\int_0^{x^{\star}} d x (\mathcal{F}_i^c- \mathcal{F}_i^d) -\int^1_{x^{\star}} d x \mathcal{F}_i^d
\end{equation}

If we now consider $x_c$, the point where the EE from the two contributions is the same, we can do the same expansion $x_c^{\star}=x_c+\phi_0^2 \delta x_c$ ($\mathcal{A}_0(x_c)=0$ by definition) :
\begin{equation}
  \phi_0^2 ( \mathcal{A}_{\Delta}(x_c)+ \mathcal{A}_{\delta x}(x_c) {\delta x_c} )=0
\end{equation}

In this way we obtain $\delta x_c$ (that only depends on $\Delta)$. In order to get $l_c$ we only have to integrate $\frac{dl}{d x}$ until $x_c$. If we expand $l_c$ then and read the contribution quadratic in the perturbation then:
\begin{eqnarray}
 \frac{l_c(\phi_0)}{(R \delta_R)^{-1} l_c(\phi_0=0)}=1+\frac{l_{\Delta}(x_c)- k_{\delta x} \mathcal{A}_{\Delta}(x_c) }{l_c(\phi_0=0)} \phi_0^2 \nonumber\\
 k_{\delta x}=\frac{\mathcal{A}_{\delta x}}{l_{\delta x}} \approx \frac{2 \left(2x_c-x_c{}^2\right){}^{3/4}}{1-x_c} \label{eq:lcrel}
\end{eqnarray}

The two integrals in $k_{\delta x}$ are divergent around $x_c$, but they diverge with the same power, so $k_{\delta x}$ is just the ratio of the coefficients that multiply the divergence. 
The coefficient $c_{l_c}$ will have contributions both from the integrals and from $\delta_R$ and will be the quantity of our interest. To compute it we only have to do two integrals: $l_{\Delta}$ and $\mathcal{A}_{\Delta}$ for every $\Delta$.

The critical width of the unperturbed system $x_c$ will be : $x_c \simeq 0.15406$.

\begin{figure}
  
    \centering
    \includegraphics[width=0.5\textwidth]{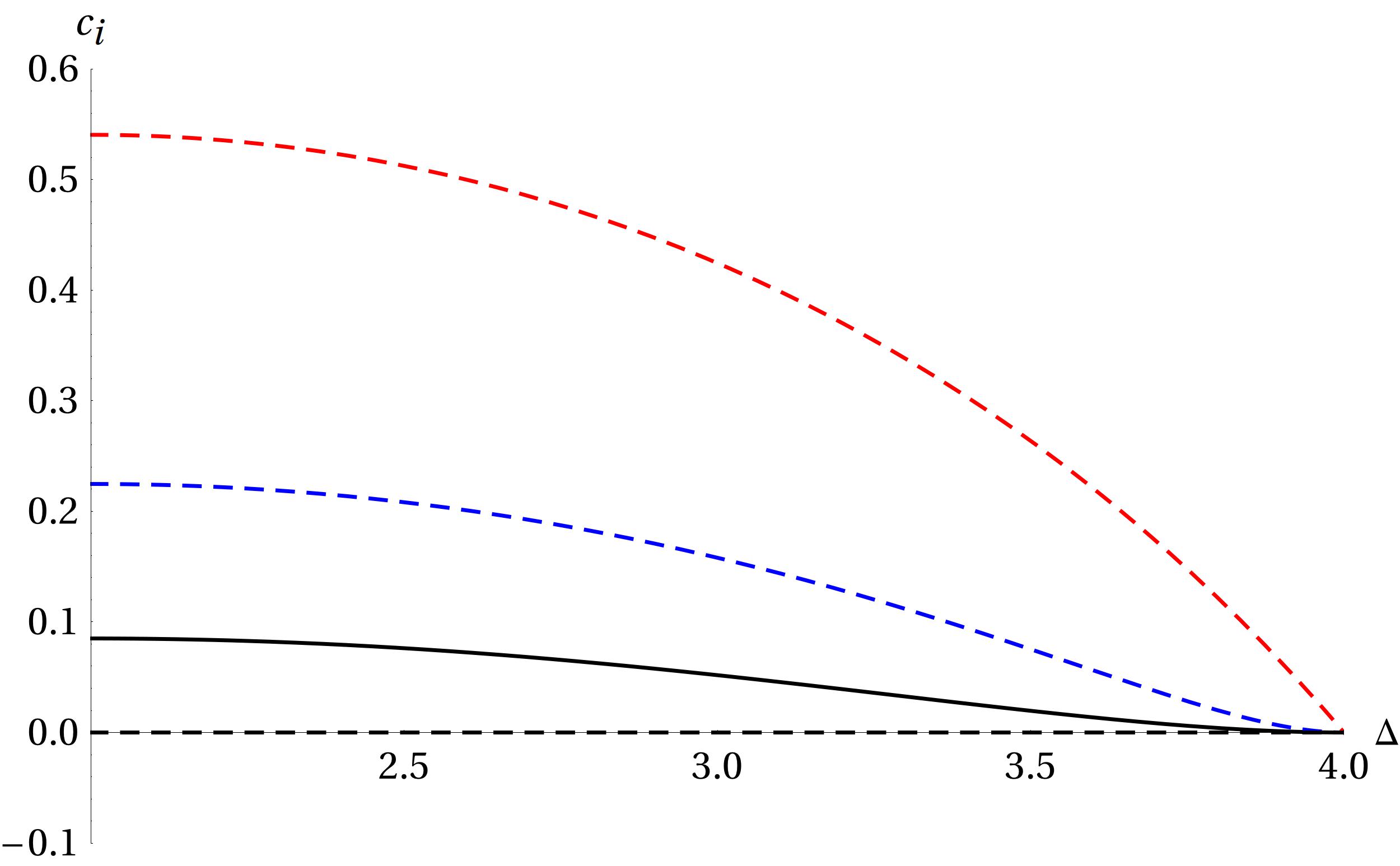}
    \includegraphics[width=0.5\textwidth]{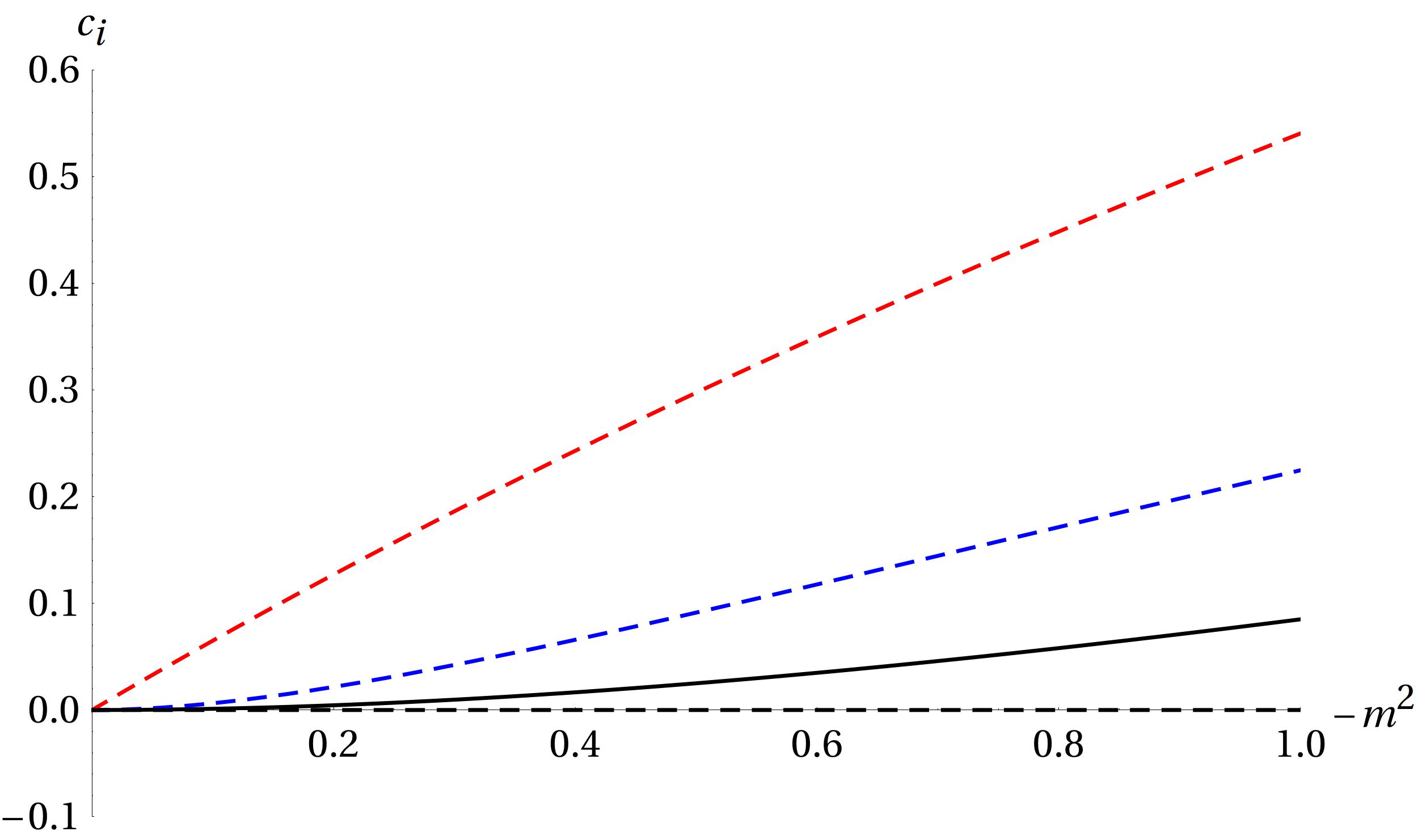}
  \caption{(Colour online) Dependence on $c_i$ on $\Delta$ and $|m^2|$ for $c_{l_c}$ (black) and the scales of our theory 
(dashed) $c_{1/T_c}=0$ (black), $-c_{\sqrt{T_{str}}}$ (red) and $c_{M^{-1}_{gb}}$ (blue) \label{fig:relevant}} 
\end{figure}

In this way, we have all we need to study how does the behavior of $l_c$ compare with that of the other physical quantities. In figure \ref{fig:relevant} we can see plotted the four $c$ of interest as a function of $\Delta$ and of  $|m^2|$ (remember that $m$ is a purely gravitational quantity and $T_c$ doesn't depend on $\Phi_0^2$).

 In this case we see how the tension doesn't seem to capture as well the behavior of $l_c$ as the other two quantities. It's worth mentioning that $c_{l_c}=-c_R+b$ (by $b$ we mean the slope of (\ref{eq:lcrel})), so $c_{l_c}$ considerably smaller than $c_R$ (is the same than the tension) requires some tuning.

\section{Discussion}
In section 2 we compared $l_c$ with three quantities that characterize the confined state: the deconfinement temperature, the tension of the string and the energy scale of the highly excited glueball states (which we called glueball mass). Overall, we have seen that although the behavior of $l_c$ is not completely the same for all the cases, there are some observations that apply to the three cases.

First, as we discussed earlier, $T_{str}$ is parametrically distinct from other scales in holographic models. But, even if $T_{str}$ is not close to $l_c$, $\frac{T_{str}(x)}{T_{str}(x=0)}$ is similar to $\frac{l_{c}(x)}{l_{c}(x=0)}$.

Second, $M_{gb}$ and $T_c$ are approximately the same (although not identical) and the way we obtained $M_{gb}$ is of course approximate. Comparing these parameters with the critical width,
$l_c$ follows very closely both scales, being roughly the same. It seems to have a slight preference for $M_{gb}$, something which makes sense since correlations in the theory are dominated by its excitations. This preference reflects the intuition that correlations are minimal beyond $l_c \sim \frac{1}{M_{gb}}$.


\acknowledgments
I am extremely grateful to R C Myers for his supervision during this project, his continuous assessments, without which this paper wouldn't have occurred, and all his suggestions regarding the manuscript. I am also thankful to A Buchel for useful discussions and for reading the manuscript.
I would like to acknowledge support from Fundaci\'{o}n Caja Madrid and the Perimeter Scholars International program. Research at Perimeter Institute is supported by the Government of
Canada through Industry Canada and by the Province of Ontario through the Ministry of Research
and Innovation.

\end{document}